\documentclass[twocolumn,superscriptaddress,floatfix]{revtex4-2}
\usepackage{eurosym}
\usepackage{graphicx,bm,amsmath,amssymb}
\usepackage{xcolor}

\def\gz{\ifmmode{Z\hskip -4.8pt Z}
    \else{\hbox{$Z\hskip -4.8pt Z$}}\fi}

\newcommand{\be}{\begin{equation}}
\newcommand{\ee}{\end{equation}}
\newcommand{\bea}{\begin{eqnarray}}
\newcommand{\eea}{\end{eqnarray}}

\begin{document}

\title{Phase diagram and topology of the XXZ chain with alternating bonds and staggered magnetic field}

\author{B. F. M\'{a}rquez}

\affiliation{Centro At\'{o}mico Bariloche and Instituto Balseiro, 8400 Bariloche, Argentina}

\author{N. Aucar Boidi}

\affiliation{Centro At\'{o}mico Bariloche and Instituto Balseiro, 8400 Bariloche, Argentina}

\author{K. Hallberg}
\affiliation{Centro At\'{o}mico Bariloche and Instituto Balseiro, 8400 Bariloche, Argentina}
\affiliation{Instituto de Nanociencia y Nanotecnolog\'{\i}a CNEA-CONICET, GAIDI,
Centro At\'{o}mico Bariloche, 8400 Bariloche, Argentina}


\author{A. A. Aligia}
\affiliation{Centro At\'{o}mico Bariloche and Instituto Balseiro, 8400 Bariloche, Argentina}
\affiliation{Instituto de Nanociencia y Nanotecnolog\'{\i}a CNEA-CONICET, GAIDI,
Centro At\'{o}mico Bariloche, 8400 Bariloche, Argentina}

\begin{abstract}
The XXZ spin-half chain has Heisenberg exchange interactions $J_z$ ($J_\perp$) in the $z$ ($x,y$) direction. The model has a transition from the spin-fluid phase for $-J_\perp < J_z < J_\perp$ to the N\'{e}el phase for $J_z > J_\perp >0$. When bond alternation $\delta$ is included, the N\'{e}el phase transitions to the dimer phase for a finite value of $\delta$. We determine the phase diagram using simple topological indicators related to the polarization of both spins. When a staggered magnetic field $B$ is included, a contour plot of these indicators as a function of 
$\delta$ and $B$ determine the amount of topological quantized spin pumping around closed circuits in the 
$(\delta,B)$ plane.

\end{abstract}



\maketitle

\section{Introduction}

\label{intro}

In recent years there has been considerable attention in topological aspects of matter \cite{Ando13,Bradlyn17,Sato17}. In particular, topological quantized charge or spin pumping can be realized in time-dependent adiabatic evolution in a closed cycle in a certain space of parameters [typically two-dimensional (2D)]. This is known as a Thouless pump \cite{Thou83,Niu84,Wang13,Citro23}.

Usually, the 2D pump cycle encloses one or more critical points at which a symmetry protected topological number jumps. Outside the critical points,
the protecting symmetry is lost allowing a continuous variation of charge or spin Berry phases, the variation of which on the cycle determines the transported amount of charge or spin. Experimentally, charge transport in Thouless pumps described by the Rice-Mele chain \cite{Naka16,Lohse16} including interactions in some cases \cite{Walter23,Viebahn23} have been realized in  chains of ultracold atoms. Different variants of the model have been discussed theoretically \cite{Asb16,Hay18,Nakag18,Stenzel19,Bertok22,Citro23,Roura23,Moreno23,Aligia23,Arg24,Tada24}.

Quantum spin pumps have been also realized experimentally \cite{Schw16} and discussed theoretically \cite{Schw16,Shin5,Meid11,Zhou14,Chen20,Aligia23,Farre24}. Different realizations of Heisenberg-like chains have been achieved with ultracold atoms \cite{Chen11,Fuku13,Schw16,Kim24} including an extremely anisotropic Heisenberg model \cite{Kim24}. The XXZ chain is the Heisenberg chain with anisotropic nearest-neighbor (NN) exchange interactions, $J_z$ in the $z$ direction and $J_\perp$ in the other two. 
While this model has been solved exactly \cite{bethe},  different variants of the model have been studied using 
field theoretical 
\cite{Kohmoto81,Totsuka98,Koba18,Furu19,Cheng23} and
numerical \cite{Nomura94,Som1,Tzeng16,Ueda20,Farre24} methods.

The phase diagram of the spin-1/2 model including next NN interactions has been calculated using the method of crossings of  excited energy levels (MCEL), supported by results of conformal field theory and the renormalization group \cite{Nomura94,Som1}. The corresponding diagram for the model with alternating NN interactions [$H_0$ in Eq. (\ref{ham})] has been calculated using different numerical techniques \cite{Tzeng16,Farre24}. 
In spite of some quantitative differences, qualitatively the results of both works coincide. 
Without alternating interactions
($\delta=0$), there is a transition at $J_z=J_\perp$ between
the spin-fluid phase for $-J_\perp < J_z < J_\perp$ to the N\'{e}el phase for large $J_z$ \cite{bethe,Nomura94,Som1}. For an infinitesimal alternation, the spin fluid phase is transformed to a dimer phase, while a finite $|\delta|$ is needed for the transition between the N\'{e}el phase and a dimer phase. The dimer phases are different for opposite signs of 
$\delta$. A similar model with experimental relevance has been studied numerically \cite{Ueda20}.

Recently, real-time dynamics of Thouless pumps in the XXZ model with alternating NN interactions including a staggered magnetic field $B$ have been calculated with the infinite time-evolving block decimation (iTEBD) method \cite{Farre24}. The authors considered different circuits in the $(\delta,B)$ plane that touch but do not cross the singular point $\delta=B=0$. The authors find that quantized pumping takes place only for $J_z > J_\perp$, when the N\'{e}el phase separates the two dimer phases for positive and negative $\delta$.

The purpose of this paper is twofold: (i) to recalculate the phase diagram of the model for $B=0$ (which we call $H_0$) using topological indicators based on simple position operators, showing the power of this method. (ii) To show how a contour plot of expectation values of the position operators in the $(\delta,B)$ plane allows the prediction of the
transported spin in pumping circuits.

The paper is organized as follows. In Sec. \ref{model} we present the Hamiltonian of the model with its known limiting cases. In Sec. \ref{methods} we define the topological indicators used to determine the phase diagram of $H_0$ and discuss some simple cases in which these indicators characterize the different phases. Sec. \ref{phdiag} presents the results for the phase diagram of $H_0$ compared with the previous results obtained by Tzeng \textit{et al.} \cite{Tzeng16}
using different sophisticated numerical methods, showing a quantitative agreement. In Sec. \ref{contour} the contour plots of the topological indicators in the 
$(\delta,B)$ plane show that these can predict the evolution of the spin transport in adiabatic Thouless pump cycles. 
At the end, a summary and discussion can be found in Sec. \ref{sum}. 

\section{Model Hamiltonian}
\label{model}

The Hamiltonian of the spin-1/2 alternating XXZ chain, 
including a staggered magnetic field reads

\begin{eqnarray}
H &=&H_{0}+H_{B},  \nonumber \\
H_{0} &=&\sum\limits_{j=1}^{N}\left[ 1+(-1)^{j}\delta \right] [
J_{\perp}(S_{j}^{x}S_{j+1}^{x}+S_{j}^{y}S_{j+1}^{y}) +\nonumber \\
&&\hspace{2.6cm}+\hspace{0.1cm}J_{z}S_{j}^{z}S_{j+1}^{z}], \nonumber \\
H_{B} &=-&\sum\limits_{j=1}^{N}(-1)^{j}BS_{j}^{z},
\label{ham}
\end{eqnarray}
where $\mathbf{S}_{j}=(S_{j}^{x},S_{j}^{y},S_{j}^{z})$ is the spin of the site $j$ of the chain. 
Since we use periodic boundary conditions,
the number of sites $N$ should be even for the system to contain an integer number of unit cells. 

The Hamiltonian for $B=0$ ($H_0$) has been studied before \cite{Tzeng16,Elben20} and can be experimentally realized with cold atoms, where the bond alternation is achieved by fine tuning the intensity of the Raman laser beams \cite{Duan3,Chen11}. 

Two limits of the model are well known. Without bond alternation ($\delta=0$) the model has been solved exactly using the Bethe ansatz \cite{bethe}. If in addition $J_z=0$, the model can be mapped to a simple fermionic tight-binding chain using a Jordan-Wigner transformation \cite{Tzeng16}. Thus this limit provides a simple picture of the spin-fluid phase, where the spins are delocalized in the chain. Instead, for large positive $J_z$ in an infinite system, there is a spontaneous symmetry breaking to a N\'{e}el or anti-N\'{e}el state ($\uparrow \downarrow \uparrow \downarrow ...$ or 
$\downarrow \uparrow \downarrow \uparrow ...$ plus fluctuations). For a finite system, the ground state is a mixture of both. The transition between the spin-fluid and the N\'{e}el phase is exactly at the isotropic point $J_z=J_\perp$ \cite{bethe,Nomura94,Som1}. The gap in the N\'{e}el phase is exponentially small near the transition and behaves as $\Delta \sim J_z- 2J_\perp$ for large $J_z$ \cite{bethe}. 

Introducing $\delta$ for $J_z=0$, the system can still be mapped onto a non-interacting fermionic chain \cite{Tzeng16,Su79}. A dimerized phase in which the expectation value of the NN singlets is larger (smaller) at the odd bonds with respect to the even ones is formed for $\delta <0$ ($\delta >0$). The odd bonds are those between sites $j$ and $j+1$ with $j$ odd. The difference between both expectation values as well as the energy gap are both proportional to $\delta$. Instead, in the isotropic case $J_z=J_\perp$, it has been shown by bosonization \cite{Cross79} that the bond-order parameter $\left\langle (\mathbf{S}_{1}-\mathbf{S}_{3})\cdot \mathbf{S}_{2}\right\rangle \sim \delta ^{1/3}$, while the energy gap $\Delta \sim \delta ^{2/3}$.

For $J_z>J_\perp$, the N\'{e}el phase competes with the dimer 
phases \cite{Tzeng16}, as described in detail in Sec. \ref{phdiag}.
Including magnetic field, pumping circuits in the 
$(\delta,B)$ plane
were studied, which are discussed in Sec. \ref{contour}.

\section{Methods}
\label{methods}

Our calculations for the model Eq.~(\ref{ham}) are based on two phases for the model defined in a ring

\begin{equation}
\alpha_s=\alpha (1,-1), \hspace{0.5cm} \alpha_{\uparrow }=\alpha (1,0), 
\end{equation}
where
\begin{equation}
\alpha (m_{_{\uparrow }},m_{\downarrow }) 
=\text{Im} \text{ ln} \left\langle U(m_{\uparrow },m_{\downarrow})\right\rangle 
\text{ mod }2\pi,  
\label{alp}
\end{equation}
is the phase of the expectation value of the following operator 
\begin{equation}
\displaystyle U(m_{_{\uparrow }},m_{\downarrow })=\exp \left[ i(2\pi /L)\sum_{j}x_{j}\left( m_{\uparrow }\hat{n}_{j\uparrow }
+m_{\downarrow }\hat{n}_{j\downarrow }\right) \right] ,  \label{uo}
\end{equation}
where $x_{j}$ is the position of the site $j$, $\hat{n}_{j\uparrow }=1/2+S_{j}^{z}$, $\hat{n}_{j\downarrow }=1/2-S_{j}^{z}$, $m_s$ (where $s=\uparrow,\downarrow$) can take the values $-1$, 0 or 1 and $L=Na/2$ is the length of the system, where $a$ is the lattice 
parameter for $\delta \neq 0$ (two times the NN distance) and $N$ is even. 

For fermionic systems it has been shown that changes in 
$\alpha(1,1)$ \cite{Resta98,Resta99,Wata18} (or more appropriately $\alpha(l,l)$ with suitably chosen $l$\cite{Aligia99,Aligia23}) are proportional to the average displacement of the particles and, in the thermodynamic limit, coincide with the charge Berry phase (the phase obtained changing the twisted boundary conditions from 0 to $2 \pi$), which in turn gives information on the polarization of the system (changes in polarization are proportional to the corresponding changes in the charge Berry phase) \cite{Ortiz94,Wata18}.
Similarly, $\alpha(1,-1)$ is an approximation to the spin Berry phase \cite{Aligia0}  which is proportional to the difference of polarizations between electrons with spin up and down \cite{gs}. Performing the Jordan-Wigner transformation mentioned in Sec. \ref{model}, it is easy to show that $\alpha_{\uparrow }$ is related to the charge Berry phase of the equivalent spinless fermionic model.
In recent years, similar expectation values been used in different cases \cite{Koba18,Het19,Furu19,Unanyan20,Deh21,Dubi21,Braver22,Tada24}.

\begin{figure}[h!]
\centering
\includegraphics[width=0.7\columnwidth]{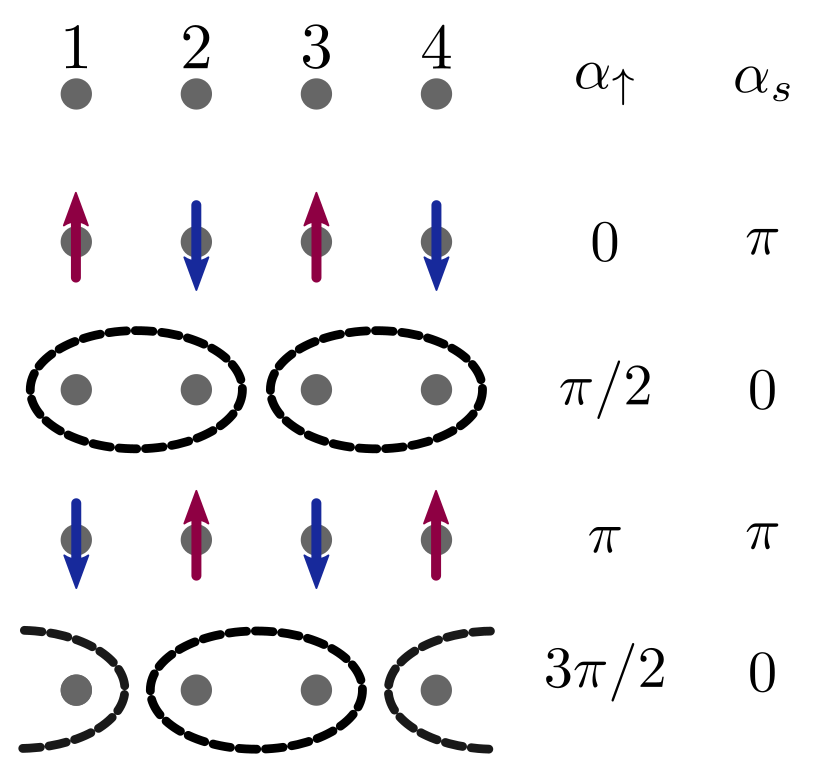}
\caption{(Color online) Schematic representation of simple states and the corresponding values of
$\alpha_\uparrow$ and $\alpha_s$. The grey circles represent the sites, arrows represent the spin degree of freedom and dashed ellipses indicate dimerization between sites.}
\label{states}
\end{figure}

To gain some intuition we discuss $\alpha_{\uparrow }$ in simple cases which are schematically represented in Fig. \ref{states}. Assume for the moment that $N$ is multiple of four and that the origin of coordinates is chosen so that $x_j=ja/2$.  Assume that the ground state is a pure N\'{e}el one with spins up occupying the odd sites (represented in the second row of Fig. \ref{states}). Then the factor entering in the exponent of Eq. (\ref{uo}) becomes $(2\pi /N)\Sigma_{j}(2j+1)=\pi N/2 \equiv 0 \text{ mod }2\pi$ and $\alpha_{\uparrow }=0$. The anti-N\'{e}el state is obtained shifting all spins half a lattice parameter ($a/2$) to the left or to the right (fourth row of Fig. \ref{states}). This amounts to replacing $2j+1$ by $2j$ or $2j+2$ above and in both cases $\alpha_{\uparrow }=\pi$. More generally, it is easy to realize that the change in 
$\alpha_{\uparrow }$ under a displacement $d$ of all 
spins up is $\Delta \alpha_{\uparrow } = (d/a) 2 \pi$. 
Therefore for a completely dimerized state in which the spins up are on average at the center of the odd bonds (third row of Fig. \ref{states}), displaced $a/4$ with respect
to the N\'{e}el state, one expects $\alpha_{\uparrow }=\pi/2$. For the other dimer state (fifth row of Fig. \ref{states}) one expects $\alpha_{\uparrow }=3\pi/2 \equiv -\pi/2$. Explicit calculations done in Ref. \cite{Aligia23} confirm these values.

Extending this argument to $\alpha_s$ is straightforward and one obtains $\alpha_s=0$ for both extremely dimerized states and $\alpha_s=\pi$ for both pure N\'{e}el states. Shifting the origin of coordinates or choosing $N$ not multiple of 4 does not affect the differences between the $\alpha$ for different states \cite{Aligia23}.

For a ring, if $\delta=0$, $H$ is invariant under the reflection $R$ that passes through two opposite sites. In addition, for any value of $\delta$, $H_0$ is invariant under the reflection $\tilde{R}$ that passes through the middle point of any two opposite NN bonds. This has important consequences on the $\alpha$. 
The reflection $\tilde{R}$ 
is equivalent to a change of sign of $x_j$ plus a translation in a NN distance $a/2$. This change of sign is equivalent to a complex conjugation in Eq. (\ref{uo}), while the translation, as explained above,   modifies the exponent of $U(1,0)$ ($U(1,-1)$) by $\pi$ ($2\pi \equiv 0$). This implies 

\begin{eqnarray}
\tilde{R}U(1,0)\tilde{R}^{\dagger }
&=&-\overline{U}(1,0), \nonumber \\
\tilde{R}U(1,-1)\tilde{R}^{\dagger }&=&\overline{U}(1,-1). 
\label{ref}
\end{eqnarray}
Since the ground state for any finite ring is unique, both $\alpha$ should remain invariant under inversion and this implies that $\alpha_{\uparrow }$ can only by either $\pi/2$ or $-\pi/2$ mod $2 \pi$, while $\alpha_s$ can only be 0 or $\pi$ mod $2 \pi$. Thus both $\alpha$ become $Z_2$ topological numbers. 

As explained above, the two extremely dimerized phases are characterized by $\alpha_{\uparrow }=\pm \pi/2$ with the sign
depending if the singlets occupy odd or even NN bonds. However, by continuity, for each dimer phase, the same topological number characterizes the whole dimer phase until a phase transition or level crossing takes place. In fact, for $|J_z| < J_\perp$ it jumps at $\delta=0$ signaling the transition between both dimer phases. This jump is related to the jump in the winding number and charge Berry phase (Zak phase for $J_z=0$) \cite{Asb16,Farre24} that takes place 
in the fermionic Su-Schrieffer-Hegger model obtained using a Jordan-Wigner transformation\cite{Tzeng16}. 

In contrast, $\alpha_{\uparrow }$ is unable to detect the transition to the N\'{e}el phase. As explained with the simple arguments above, the extreme N\'{e}el and anti-N\'{e}el states have $\alpha_{\uparrow }=$ 0 or $\pi$. For a finite system the unique ground state contains a measure of both states and $\alpha_{\uparrow }= \pm \pi/2$ as required by the topological protection by the reflection symmetry. Instead, as also discussed above, $\alpha_s=0$ for both extreme dimer phases and $\alpha_s=\pi$ for both pure N\'{e}el states. By continuity, the same values extend to the whole dimer and N\'{e}el phases and therefore the jump in $\alpha_s$ denotes the dimer-N\'{e}el phase transition. In fact, in the thermodynamic limit, this jump coincides with the jump in the spin Berry phase \cite{Aligia0}. In turn, the latter jump corresponds to a crossing of excited energy levels used by the MCEL to determine the transition \cite{Nomura94,Som1}. 

The MCEL coincides with jumps of the Berry phases also in other models \cite{Torio1}. Furthermore, it has been shown that for a general extended Hubbard model including NN repulsion, density-dependent hopping, pair hopping and exchange, in a wide range of parameters with weak interactions,
for which continuum-limit field theory techniques
are expected to be quantitatively reliable, 
the bosonization results for the phase diagram coincide with those obtained numerically from jumps in charge
and spin Berry phases \cite{AA99}.

The key point for what follows is that 
$\alpha_{\uparrow }$ and $\alpha_{s}$ complement each other and characterize the three different topological phases of $H_0$. In Sec. \ref{phdiag} we determine the phase diagram of $H_0$ using these topological indicators.

When a staggered magnetic field $B$ is included, the reflection symmetry $\tilde{R}$ is broken and the $\alpha$ lose their topological protection for $\delta \neq 0$, but they continue to give information on the change in the position of the spins under changes in the parameters. In particular, a contour plot of them in a two-dimensional parameter space, allows us to predict these changes, under an 
adiabatic cycle. This is shown in Sec. \ref{contour}.

For $\delta=0$, the reflection $R$ is a symmetry of the Hamiltonian for any $B$, and an argument similar but simpler  to that used to analyze the consequences of 
$\tilde{R}$, shows that both $\alpha_{\uparrow }$ and 
$\alpha_{s}$ can only be either 0 or $\pi$ mod $2 \pi$.

At this point, we explain why we use periodic boundary
conditions (PBC). 
When open boundary conditions (OBC) are used, 
in general both reflection symmetries $R$ and 
$\tilde{R}$ are lost. If either $\delta $ or $B$ are different from zero, the unit cell contains two sites and in order to have an integer number of unit cells, the number of sites $N$ should be
odd. Therefore the only possible reflection symmetry 
is $R$ through the site $(N+1)/2$ at the middle of the chain. However the exchange interactions are different at the left and the right of this point. 
Therefore, the topological protection is lost.
One might expect that for a long enough chain 
similar results are obtained. However for the charge transition in the ionic Hubbard model, results 
for the charge gap near the charge transition with 
$N \sim 30$ using PBC seem superior to  $N \sim 400$
using OBC \cite{Moreno23}. Also, as discussed 
at the beginning of the next Section, 
for our system $\alpha_s$ becomes smooth 
and does not jump for $N \sim 60$
using OBC \cite{Farre24}. 

While the information above is enough to understand
the results presented below, in the rest of this Section
we discuss more technical points, in particular in relation with continuum-limit field theory. Using bosonization techniques, it has been shown that the 
operator $U(m_{\uparrow },m_{\downarrow })$ takes 
the form

\begin{equation}
U=\exp \left[ -i c \phi_a \right] \text{,~~~ } 
\phi_a= \frac{1}{L}\int dx \phi(x),    
\label{ubos}
\end{equation}
where $\phi(x)$ is a charge or spin field, $\phi_a$ its 
average and $c$, like $\phi(x)$, depends on the
particular $m_{\sigma}$ \cite{Ali05,note}. A similar expression that differs in an irrelevant constant
was derived more recently \cite{Furu19}. 

The scaling 
of the expectation value of $U$ in gapless phases 
of the XXZ chain \cite{Koba18,Furu19} and ’t Hooft anomalies \cite{Cheng23} of the model have been studied using conformal field theory and renormalization group (RG). In addition, similar models were studied 
with RG \cite{Kohmoto81,Totsuka98}.

\section{Phase diagram of $H_0$}
\label{phdiag} 

The phase diagram of $H_0$ has been calculated by Tzeng 
\textit{et al.} using R\'{e}nyi entropies and the second derivative of the ground state energy obtained with density matrix renormalization group (DMRG) in systems with up to $N \sim 120$ \cite{Tzeng16}. Recently a calculation of the phase diagram also using DMRG 
with $N \sim 60$ has been reported \cite{Farre24}.  

From these works, as well as the jump of 
$\alpha_{\uparrow }$, it is clear that the boundary between both dimer phases for $|J_z|\leq J_\perp$ is at 
$\delta=0$. In Ref. \cite{Farre24}, a calculation
of the Zak Berry phase $\varphi$ (which in principle should
give the same information as $\alpha_{\uparrow }$ 
except for a constant \cite{Aligia23,Tada24}) is reported.
Instead of a jump in $\pi$ at $\delta=0$, $\varphi$ shows
a continuous evolution with a total change of $\pi$
as $\delta$ is varied. This is likely due to the use of
open boundary conditions in Ref. \cite{Farre24},
under which the protecting symmetry $\tilde{R}$ is lost 
[see Eq. (\ref{ref})].

For $J_z > J_\perp$, the N\'{e}el phase appears at 
$\delta=0$ and the boundary between this phase and any of the dimer phases is not trivial to determine. 
In particular, there is a quantitative discrepancy 
between Refs. \cite{Tzeng16} and \cite{Farre24} for small
$J_z - J_\perp$. The extent of the N\'{e}el phase is larger in the latter.

Here we use $\alpha_s$ as a topological indicator to detect the transition, extrapolating the value of 
$\delta$ at which $\alpha_s$ jumps in rings of $N$ sites with even $N$ in the range $12 \leq N \leq 24$ using exact diagonalization.  For $J_z < 1.5 J_\perp$ we also include $N=26$.  
The extrapolation was done using a quadratic function in $1/N$, as in Ref. \cite{Moreno23}. To estimate the error, we repeated the extrapolation taking out the point with largest $N$. The difference between both values of $\delta$ is always less than $1.5 \times 10^{-3}$ and, in general, less than $5 \times 10^{-4}$ for $J_z/J_\perp \geq 2$.

\begin{figure}[h!]
\begin{center}
\includegraphics[width=\columnwidth]{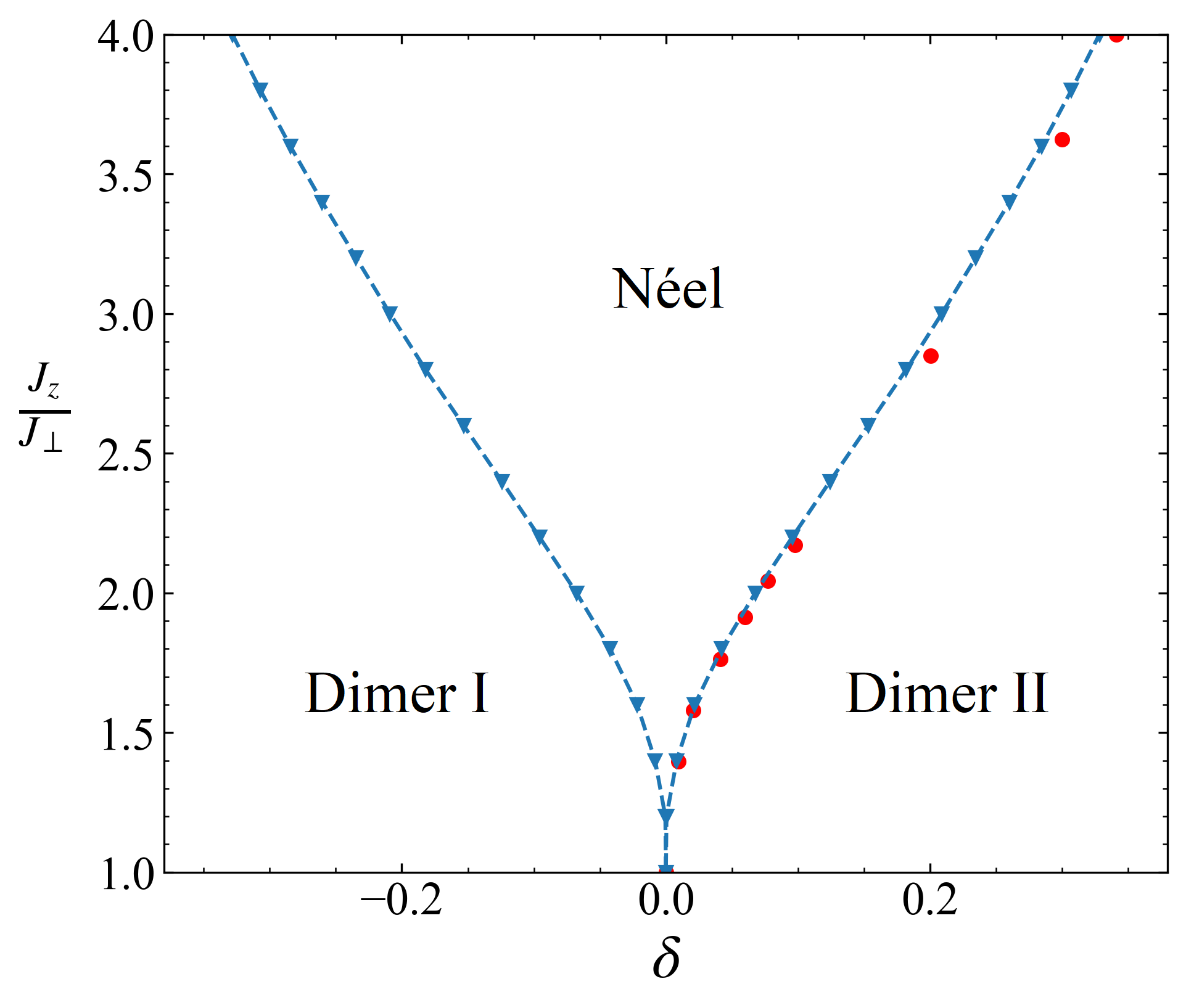}
\caption{(Color online) Phase diagram of $H_0$ obtained from the jump $\alpha_s$ (blue triangles) and compared with previous results (red circles) \cite{Tzeng16}. Dimer I (II) corresponds to larger
$\langle \mathbf{S}_{j} \cdot \mathbf{S}_{j+1} \rangle$ in the odd (even) bonds.}
\label{diag}
\end{center}
\end{figure}

The resulting boundaries of the N\'{e}el phase are shown
in Fig. \ref{diag}, and compared with previous ones obtained using finite-size scaling of the R\'{e}nyi entropies $S_2$ and the second derivative of the ground state energy, obtained with DMRG in systems with up to $N \sim 120$ sites \cite{Tzeng16}. The agreement is noticeable. This shows the power of our method which is computationally much less expensive. However, we believe that the results so far are not accurate enough in the region near $J_z/J_\perp=1$.

A characteristic of the phase diagram is that the region of the N\'{e}el phase is very narrow for small $J_z/J_\perp - 1$. This is expected, since at the point $J_z/J_\perp - 1=\delta=0$, the effect of a small increase in $J_z$ opens a gap exponentially \cite{bethe}, while an increase in $\delta$ opens a gap proportional to $\delta ^{2/3}$ \cite{Cross79}. However, the detailed behavior of the boundary in this zone is difficult to predict. Our value for the transition at $J_z/J_\perp=1.2$, $|\delta| \sim 5\times 10^{-4}$ is less than the above estimated error $1.5 \times 10^{-3}$. Instead, the value that we obtain for $J_z/J_\perp=1.4$, $|\delta|= 8.17 \times 10^{-3}$ (slightly smaller than that of Tzeng \textit{et al.}) is more reliable.

\begin{figure}[hb]
\begin{center}
\includegraphics[width=0.9\columnwidth]{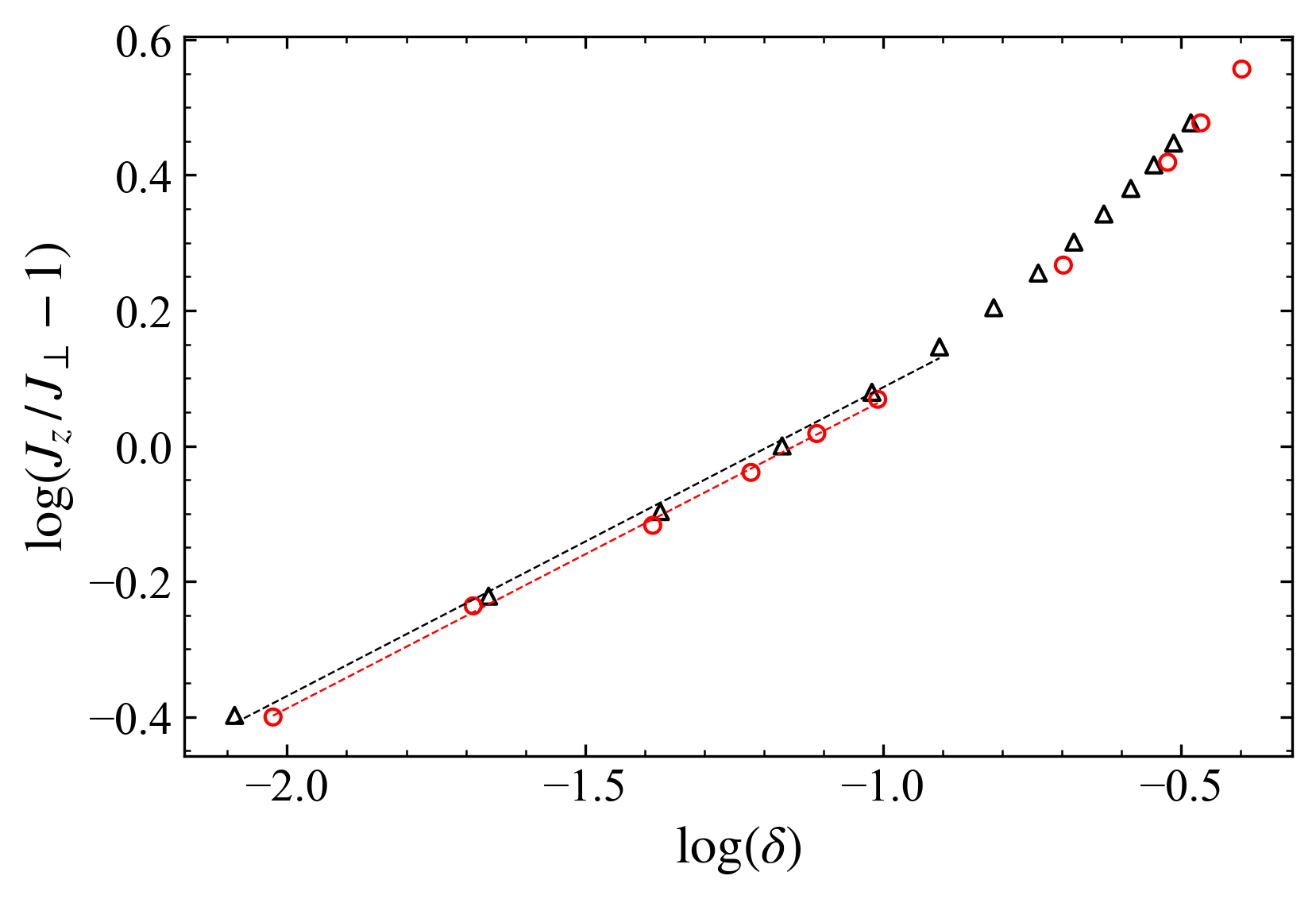}
\caption{(Color online) Critical value of $J_z/J_\perp -1$ at the
N\'{e}el-dimer transition as a function of $\delta$ (black triangles) and compared with previous results (red circles) \cite{Tzeng16}. }
\label{loglog}
\end{center}
\end{figure}

In Fig. \ref{loglog} we display the results of Fig. \ref{diag} for $J_z /J_\perp\geq 1.4$ and $\delta > 0$ using a log$_{10}$ scale for both axis. Our results and those of Tzeng \textit{et al.} \cite{Tzeng16} suggest that approximately the boundary for $1.4 \leq J_z/J_\perp \leq 2.4$ ($0.0082 \leq \delta \leq 0.124$)  corresponds to $J_z/J_\perp -1\sim \delta^{0.46}$. Instead,
the results of Ref. \cite{Farre24} suggest $J_z/J_\perp -1\sim \delta$. 

\section{Contour plots of the phases}
\label{contour}

\begin{figure}[th]
\begin{center}
\includegraphics[width=0.9\columnwidth]{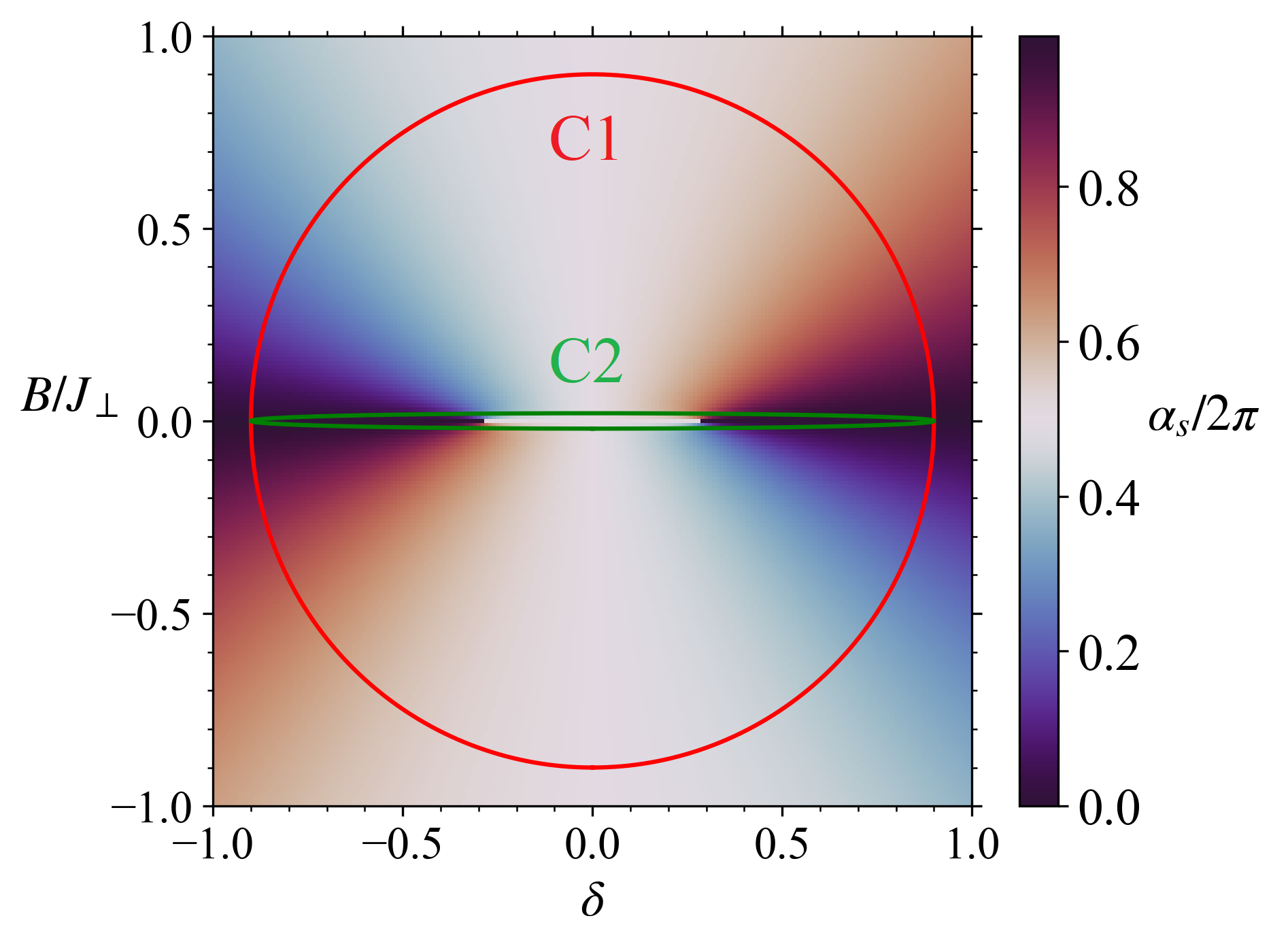}\\
\vspace{-0.5cm}
\includegraphics[width=0.9\columnwidth]{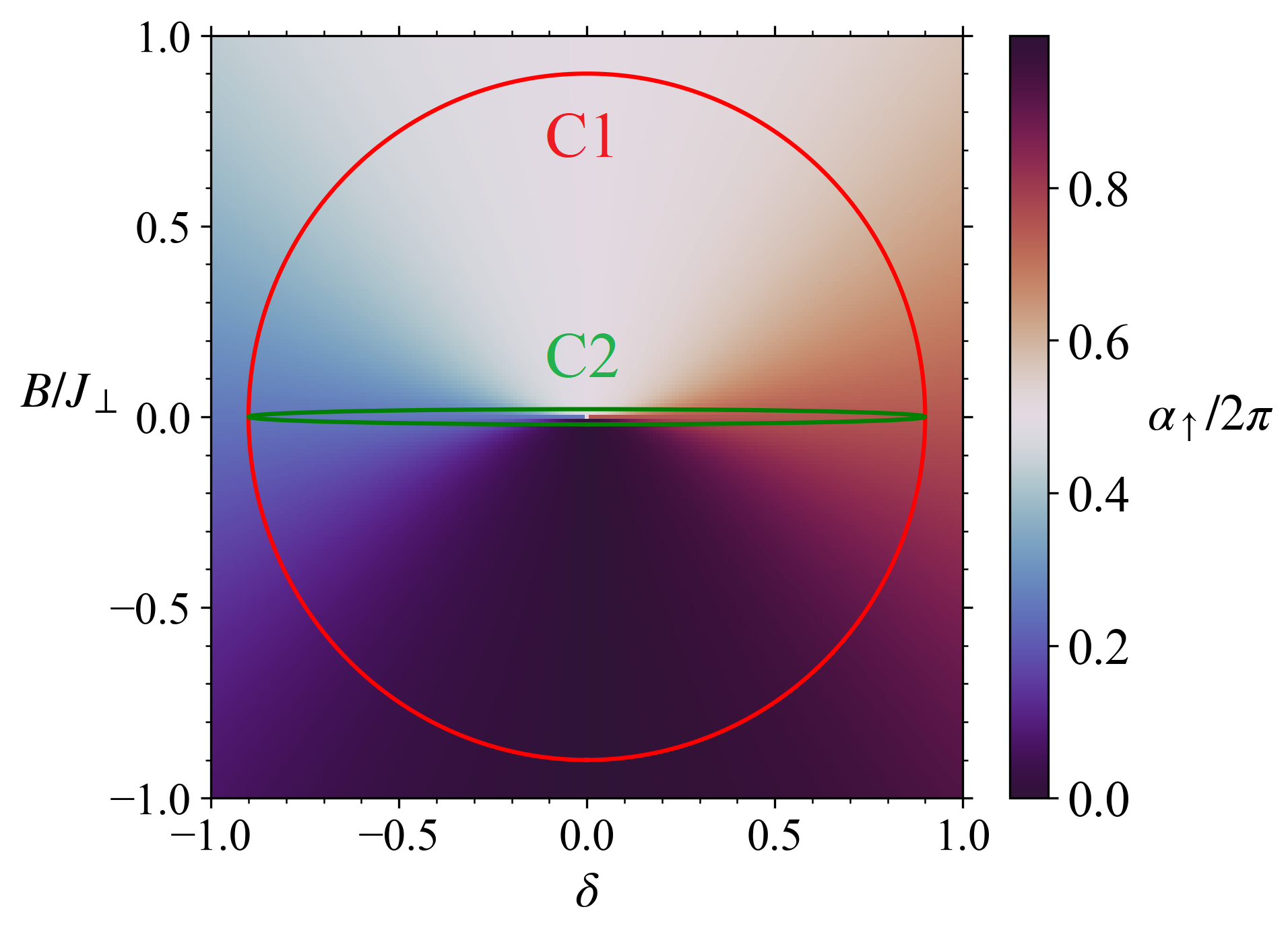}
\end{center}
\caption{(Color online) $\alpha_{s}$ (top)
and $\alpha_{\uparrow }$ (bottom) as a function of dimerization
parameter $\delta$ and staggered magnetic field for $J_z/J_\perp=2$. C1 and C2 curves represent the pumping cycles used. }
\label{conj2}
\end{figure}

In Fig. \ref{conj2} we show a contour plot of both $\alpha$ for $J_z > J_\perp$. The sharp transitions between the N\'{e}el and the dimer phases are evident in $\alpha_{s}$ but imperceptible in $\alpha_{\uparrow }$. To understand this behavior, let us analyze the change in both position indicators as the system changes slowly from the point $(\delta,B/J_\perp)=(-1,0)$ to $(0,B/J_\perp)$ with small positive $B$. The first point corresponds to a perfectly dimerized phase with singlets at the odd bonds. As discussed in Sec. \ref{methods}, this corresponds to $\alpha_{\uparrow }=\pi/2$. Since the down spins in this phase are localized at the same places as the up spins, their difference vanishes and therefore $\alpha_{s}=0$. As $\delta$ increases beyond the transition to the N\'{e}el phase, the ground state is a mixture of the N\'{e}el and anti-N\'{e}el states for small $B$. As a consequence, the spin up continues to be mostly centered at the odd bonds. Near the critical value of $\delta$, only for large positive $B$ the up spins move to the right (increasing $x_j$) occupying the even sites forming an anti-N\'{e}el state with $\alpha_{\uparrow }=\pi$. Instead, $\alpha_{s}$ jumps abruptly at the transition for $B=0$, since the {\em difference} between positions of up and down electrons moves from 0 to the NN distance as the system enters the N\'{e}el phase. When $\delta=0$ is reached even for small positive $B$ the anti-N\'{e}el phase is favored with $\alpha_{\uparrow }=\alpha_{s}=\pi$. 

Summarizing, in the displacement from $(\delta,B/J_\perp)=(-1,0)$ to $(0,0^+)$, one starts with $\alpha_{\uparrow }=\pi/2$ and $\alpha_{s}=0$ reflecting the fact that both spins are centered at the even bonds, and at the end $\alpha_{\uparrow }$ ($\alpha_{s}$) has increased by $\pi/2$ ($\pi$) as a consequence of the displacement of the spins up (down) by a quarter of a unit cell to the right (left).

One can continue the evolution to a point $(0.6,0)$ for example to the other dimerized phase, implying a displacement of the spins another quarter of a unit cell in the same direction, reaching the values $\alpha_{\uparrow }=3 \pi/2$ and $\alpha_{s}=2 \pi \equiv 0$ mod $2 \pi$. Finally, one can close the circuit with any path with $B<0$ with a total change of $2 \pi$ in $\alpha_{\uparrow }$, which means a quantized transport of a spin up to the next unit cell to the right, and a total change of $\alpha_{s}$ in $4 \pi$ indicating that the spins down moved in the opposite direction. In fact, all closed circuits that enclose once the critical segment between critical points $(-\delta_c,0)$ and $(\delta_c,0)$, without crossing it, in the same direction, are topologically equivalent and lead to the same spin transport in an adiabatic time cycle. This is also the case for a circuit studied recently in which this critical segment is touched but not crossed \cite{Farre24}.

\begin{figure}[th]
\begin{center}
\includegraphics[width=\columnwidth]{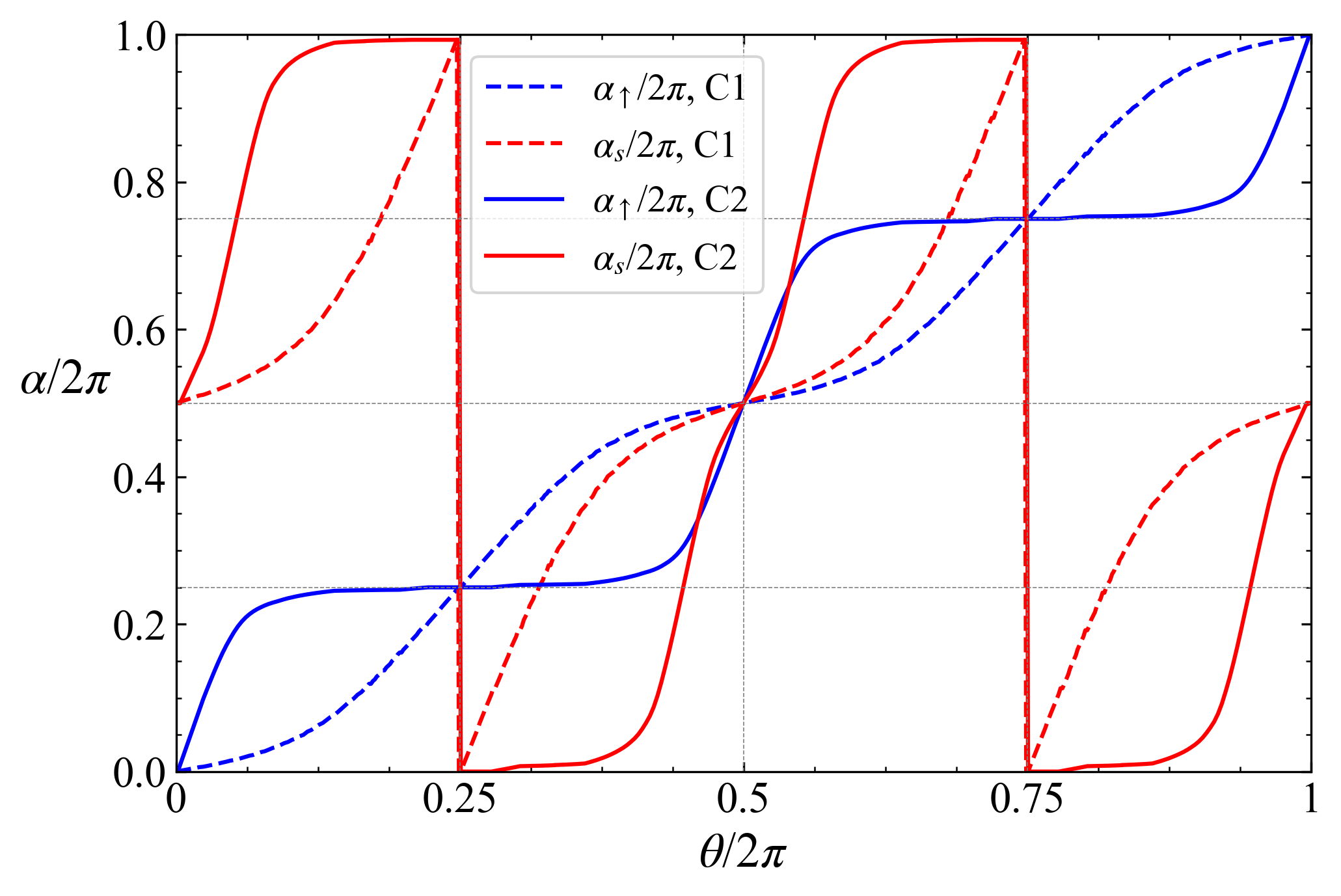}
\end{center}
\caption{(Color online) $\alpha_{\uparrow}$ and $\alpha_s$ as a function of the adiabatic parameter $\theta$ through the pumping circuits defined by Eqs. (\ref{pumping}) for $J_{z}/J_\perp=2$. Dashed (full) lines correspond to C1 (C2). }
\label{pumping_fig}
\end{figure}

As examples, in Fig. \ref{pumping_fig} we show the change in $\alpha_s$ and $\alpha_{\uparrow}$ for pumping circuits defined by

\begin{equation}
\begin{split}
\delta(\theta)&=-0.9 \sin{(\theta)}\\
B(\theta)&=-B_0 \cos{(\theta)},
\label{pumping}
\end{split}
\end{equation}

\noindent as $\theta$ is increased adiabatically from 0 to $2 \pi$ closing the circuit. We have chosen $B_0/J_\perp=0.9$ (0.02) for the circuits C1 (C2) indicated in Fig. \ref{conj2}.

For the circuit C1, the indicators change smoothly passing through topologically protected points when $\theta$ is multiple of $\pi/2$. Specifically for $\theta=0$, the system is at the point $(\delta,B/J_\perp)=(0,-0.9)$, where the ground state is the N\'{e}el state plus fluctuations, with $\alpha_{\uparrow }=0$ and $\alpha_{s}=\pi$, as described in Sec. \ref{methods}. Increasing successively $\theta$ in $\pi/2$, the changes in thermodynamic phase and the values of $\alpha_{\uparrow }$ correspond to those represented in Fig. \ref{states} jumping to the next row: odd dimers with $\alpha_{\uparrow }=\pi/2$ at $(-0.9,0)$, anti-N\'{e}el with $\alpha_{\uparrow }=\pi$ at $(0,0.9)$ and even dimers with $\alpha_{\uparrow }=\pi/2$ at $(0.9,0)$. At the end of the cycle a spin up has been transported to the next unit cell, corresponding to a total change in $\alpha_{\uparrow }$ by $2 \pi$. The spin down is displaced in the opposite direction with a total change of $\alpha_{s}$ by $4 \pi$.

For the circuit C2, the total spin transported and the values of both $\alpha$ at the topologically protected points are the same as for C1. There are however, important differences at the intermediate points. The small value of $B_0/J_\perp$ implies that moderate values of $|\delta|$ dimerize the system more easily and this fact is reflected in the values of both $\alpha$. In addition, there is a large slope in $\alpha_{s}$ at the points where $\alpha_{s}=\pi/2$ and $\alpha_{s}=3\pi/2$. This is reminiscent of the jump in $\alpha_{s}$ at the N\'{e}el-dimer transition for $B=0$ between $\alpha_{s}=\pi$ and $\alpha_{s}=2\pi \equiv 0$ mod $2 \pi$.

\begin{figure}[th]
\begin{center}
\includegraphics[width=0.9\columnwidth]{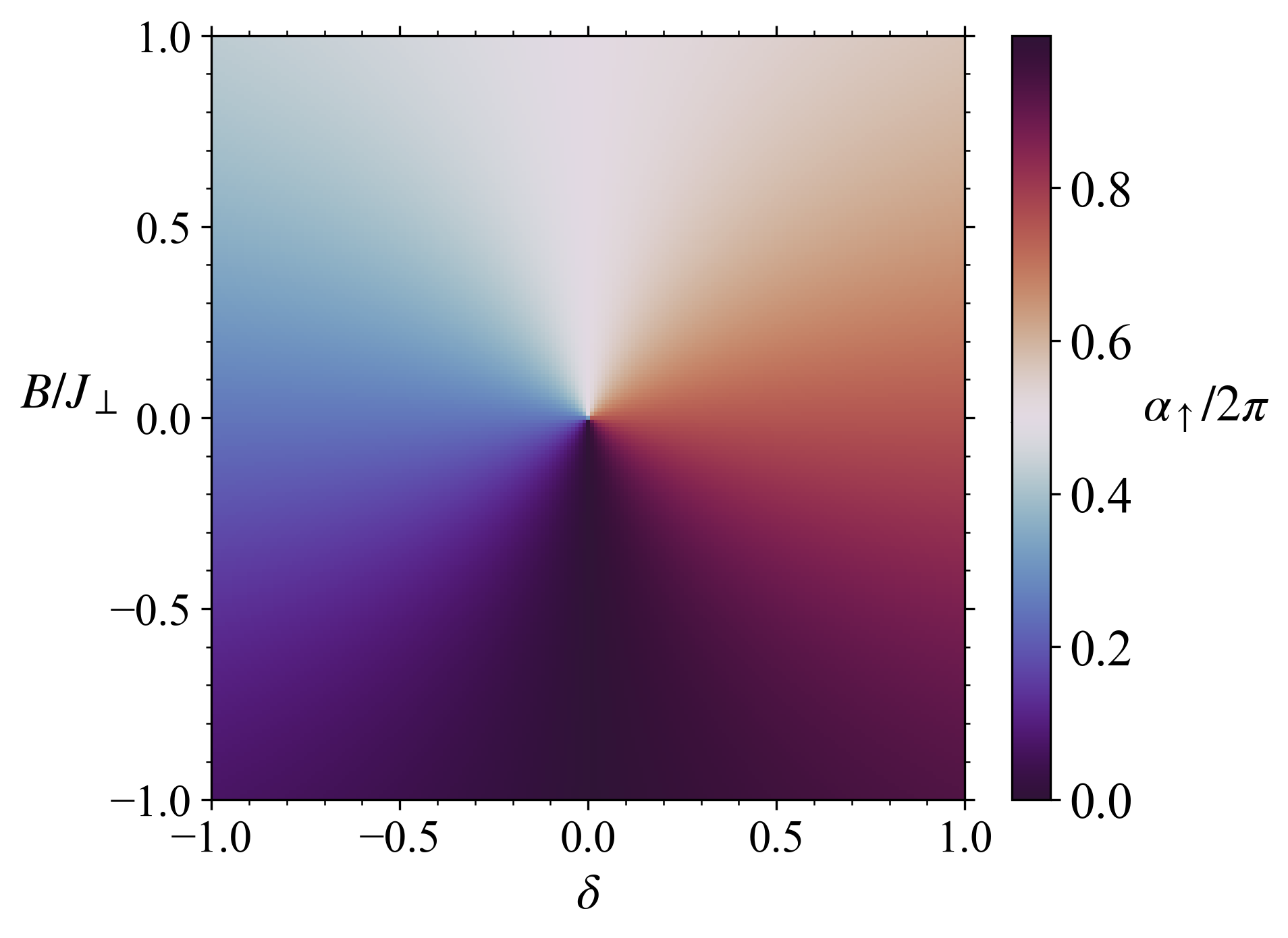}
\end{center}
\caption{(Color online)  $\alpha_{\uparrow }$ as a function of dimerization
parameter $\delta$ and staggered magnetic field for $J_{z}=0$.}
\label{conj0}
\end{figure}

For $J_z < J_\perp$, the contour plots are similar, but the segment between critical points collapses to the origin. An example is shown in Fig. \ref{conj0} for $\alpha_{\uparrow }$, which gives more information than $\alpha_s$ because the former is able to distinguish between both dimer phases and between the N\'{e}el and anti-N\'{e}el states. Adiabatic pump cycles that enclose the origin have the same topological properties as the ones discussed above, and again the contour plot allows to predict the spin transport along a time dependent path. 

However, a time-dependent calculation in a circuit that touches but does not cross the origin has shown recently that quantized spin transport is lost in this case \cite{Farre24}. We believe that the difference is that for $J_z < J_\perp$, the gap is closed at the origin, and pumping with finite velocity necessary leads to spin excitations losing the adiabatic condition. In fact, oscillations in the transported spin indicate that this is the case. In contrast in the N\'{e}el phase, the N\'{e}el and anti-N\'{e}el states are separated from the rest by a finite gap \cite{bethe}. One might wonder if the quasi degeneracy between these two states (degeneracy in the thermodynamic limit) could also affect quantized pumping. However, the mixture between both states decreases exponentially with system size, and one can expect that even crossing the segment between critical points $(-\delta_c,0)$ and $(\delta_c,0)$ by a small amount does not spoil the quantized pumping. Time-dependent calculations in the interacting Rice-Mele model also obtained that the addition of an Ising term ($J_z$) to the model stabilizes pumping (of charge in this case) despite the quasi degeneracy of the ground state \cite{Bertok22}.

\section{Summary and discussion}
\label{sum} 

We have used two position indicators $\alpha_{\uparrow }$ and $\alpha_{s}$,
to study the XXZ ring with interactions with alternation proportional to
$\delta$ and staggered magnetic field $B$. The value of 
$\alpha_{\uparrow }$
provides information on the shift in the position of the spins $\uparrow $
under changes in the parameters, while $\alpha_{s}$ accounts for the difference
in displacements between both spins.

For either $\delta=0$ or $B=0$, both indicators become topological $Z_2$ numbers protected by a reflection symmetry. For $B=0$ we have used $\alpha_{s}$ to calculate the phase diagram of the system, obtaining results that agree quantitatively with 
those of one of the groups which calculated the diagram before \cite{Tzeng16}, but with a smaller computational cost. The topological indicators $\alpha_{\uparrow }$ and $\alpha_{s}$ provide complementary information on the thermodynamic phases of the model.

We have calculated contour plots of both 
$\alpha_{\uparrow }$ and $\alpha_{s}$ in the $(\delta,B/J_\perp)$ plane, which permit to predict the evolution of the spin transport in adiabatic Thouless pump cycles. The transported spin is quantized in the cycle and only the intermediate values depend on the specific cycle for topologically equivalent cycles. Again $\alpha_{\uparrow }$ and $\alpha_{s}$ provide complementary information. Since experimentally the occupation numbers in different sites are measured simultaneously \cite{Schw16}, it is in principle possible to observe differences in the behavior of $\alpha_{\uparrow }$ and $\alpha_{s}$ for pumping circuits that pass near critical points and for small number of particles, for which the entanglement is larger. 

\section*{Acknowledgments}
We thank Daniel Cabra for helpful discussions.
AAA and KH acknowledges financial support provided by PICT 2020A 03661 and PICT 2018-01546 of the ANPCyT, Argentina. 

\bibliography{marquez_xxza12.bib}

\begin{thebibliography}{60}
\bibitem [{not()}]{note}%
  \BibitemOpen
  \href@noop {} {\ }\bibinfo {note} {{There is a mistake in the sign of the exponent in Ref. \cite{Ali05} because the shift in $\Pi_c(x)$ is positive}}\BibitemShut {NoStop}%
\end{thebibliography}

\end{document}